# Casimir force in discrete scalar fields I: 1D and 2D cases

Eduardo Flores, Christian Ireland, Nabil Jamhour, Victor Lasasso, Nicholas Kurth, & Matthew Leinbach

*Department of Physics & Astronomy*
*Rowan University*
*Glassboro, NJ 08028*

We calculate the Casimir force between parallel plates for a massless scalar field. When adding the energy of normal modes, we avoid infinities by using a discrete spacetime lattice; however, this approach proves ineffective as long as both space and time are kept discrete. Yet, when time is treated as continuous while the scalar field forms a spatial periodic lattice, our method succeeds, and we refer to this approach as Hamiltonian lattice theory. The dispersion relation for both square and triangular lattices accurately reproduces the subtle Casimir effect, providing evidence that the Casimir force is independent of the type of lattice used. At low frequencies, both lattices exhibit a high level of rotational symmetry. However, at high frequencies, they lose this symmetry, even though the propagation of high-frequency waves becomes limited as their group velocity approaches zero.

## Introduction

Quantum field theory predicts that the vacuum is filled with zero-point energy. In continuous spacetime, fields have an infinite number of modes that contribute to the zero-point energy, leading to a divergent total zero-point energy. This divergence is one of the many infinite quantities in the theory. It is well-known that these infinities arise from our limited understanding of the short-distance behavior of the theory [1]. Fortunately, in the Standard Model, these infinities can be subtracted. The technique used to obtain finite results that can be experimentally tested is known as renormalization [2].

Among the calculations that require regularization to address the infinities arising in continuous field theory, the Casimir force stands as one of them. The Casimir effect was discovered and calculated long ago [3], and it has been experimentally measured by G. Bressi et al. [4]. In this work, our aim is to obtain the Casimir force without encountering infinite quantities. One approach is to employ lattice regularization, which involves temporarily discretizing spacetime and subsequently returning to continuous spacetime at the end of the calculation. While a calculation of the Casimir effect using lattice regularization has been performed before, it has been limited to certain cases [5,6].

Here, we first calculate the Casimir force for a scalar field by considering spacetime as a discrete lattice. In the 1D case, we find that the Casimir force is zero; this result is not in agreement with the standard value [7]. In 2D and 3D, this technique does not produce the Casimir force either. However, we discover that if time is kept continuous and the scalar field is treated as a spatially discrete lattice, we obtain the correct results in 1D, 2D and 3D. Our results are comparable to those obtained using dimensional regularization for infinite parallel plates embedded in a D-dimensional space, separated by a distance of 2a, and satisfying Dirichlet or Neumann boundary conditions,

$$\mathcal{F} = -\hbar c \frac{D}{\pi^{(D+1)/2} 2^{2D+2} a^{D+1}} \Gamma\left(\frac{D+1}{2}\right) \zeta(D+1), \quad (1)$$

where $\Gamma$ and $\zeta$ are the gamma and zeta functions [7].

## 1D Scalar Field

First, we calculate the Casimir force for a massless scalar field in discrete spacetime. The system consists of field points separated by distance $d$, Planck length. The size of the lattice is finite, $l$. The relation between $l$ and the number of points, $N$, is $l = Nd$. The corresponding unit of discrete time, $\tau$, satisfies the relation $d = c\tau$, where $c$ is the speed of light. The discretized wave equation for field $\phi_{j,s}$, where $j$ and $s$ are integers that correspond to space and time respectively, is

$$\frac{1}{\tau^2}(\phi_{j,s+1} - 2\phi_{j,s} + \phi_{j,s-1}) = \frac{c^2}{d^2}(\phi_{j-1,s} - 2\phi_{j,s} + \phi_{j+1,s}). \quad (2)$$

A solution to Eq. 2 is given by $\phi_{j,s} = a e^{i(kx-\omega t+\delta)}$, where $a$ is amplitude, $x = jd$ is position, $t = s\tau$ is time, and $\delta$ is a phase. Replacing the solution into Eq. 2, we obtain

$$\omega = ck, \quad (3)$$

an expression similar to the continuous case. Once boundary conditions, $\phi_{j=0,s} = \phi_{j=N+1,s} = 0$, suitable for the Casimir effect calculation are applied, the field takes the form

$$\phi_{j,s;r} = a \sin(k_r d j) e^{i\omega_r s\tau}, \quad (4)$$

where

$$\omega_r(N) = ck_r = \frac{c\pi}{(N+1)d} r, \quad (5)$$

and $r = 1, 2, \ldots, N$.

The zero-point energy in a region of length $x = nd$ is

$$U(n) = \sum_{r=1}^{n} \frac{1}{2} \hbar \omega_r. \quad (6)$$

Using Eq. 5 and simplifying we have

$$U(n) = \frac{\pi \hbar c}{4d} n \quad (7)$$

In the typical Casimir effect, a pair of conducting plates are positioned facing one another at a small distance. The Casimir force requires interaction not only between the plates, but also with the outside environment. To simulate this mathematically, we follow Cooke [8] and take a lattice of length $l$, with fix plates at the ends, and divide it into two by placing a third plate that can move along the $x$-axis, see Fig.1. We require that the field at a plate is zero. The space between the movable plate and its nearest plate is regarded as the gap width $(x)$, which represents the space between the conducting plates. The movable plate and the farther end plate model the environment, $(l - x)$.

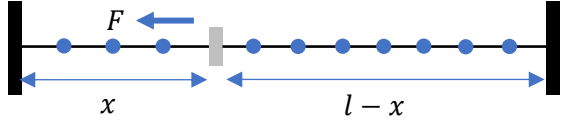

**Fig. 1** Discrete scalar field bounded by end plates. The movable plate at distance $x$ experiences an attractive force $F$ towards the nearest plate.

The gap width, $x$, is

$$x = (n+1)d \sim nd, \quad (8)$$

where $n$ is an integer and $d$ is the distance between two neighbor field points. The environment size is

$$l - x = (N - n)d. \quad (9)$$

The zero-point energy of the system is

$$U_S(n, N) = U(n) + U(N - n). \quad (10)$$

Using, Eq. 7, we get

$$U_S(n, N) = \frac{\pi \hbar c}{4d} N, \quad (11)$$

Since $U_S$ is independent of $n$ we get that the Casimir force is zero,

$$\mathcal{F} = -\frac{dU_S}{dn} \frac{dn}{dx} = 0. \quad (12)$$

This result is not in agreement with the standard value in Eq. 1. However, there appears to be alternative approaches to calculate the Casimir force using a discrete spacetime lattice [5,6].

Another approach is by starting with a discrete spacetime, then recovering the continuous limit and only then calculating Casimir force. However, we must employ a regularization technique to discard infinite values. We let $N, j, s \to \infty$ and $d, \tau \to 0$ while keeping finite the length of the lattice $l = (N+1)d$, time $t = s\tau$, and distance $x = jd$. At this limit Eqs. 4 and 5 become

$$\phi(x, t; r) = a \sin(k_r x) e^{i \omega_r t}, \quad (13)$$

and

$$\omega_r = \frac{c\pi}{l} r, \quad (14)$$

where $r = 1, 2, \ldots, \infty$. The zero-point energy in the gap of length $x$ is given by

$$U(x) = \frac{\pi \hbar c}{2x} \sum_{r=1}^{\infty} r, \quad (15)$$

The zero-point energy of the system is

$$U(x, l) = U(x) + U(l - x). \quad (16)$$

In the limit, $l \gg x$, we obtain an infinite force

$$\mathcal{F} = -\frac{dU}{dx} = \frac{\pi \hbar c}{2x^2} \sum_{r=1}^{\infty} r. \quad (17)$$

The sum may be regulated using the zeta function,

$$1 + 2 + 3 + \cdots \to -\frac{1}{12}. \quad (18)$$

From Eqs. 17 and 18 we obtain

$$\mathcal{F} = -\frac{\pi}{24} \frac{\hbar c}{x^2}. \quad (19)$$

This is similar to the Casimir force in Eq. 1 for $D = 1$ and $x = 2a$.

It is possible to keep time continuous and let only space be discrete. We find that this approach actually works for this calculation but it breaks space-time symmetry. The discrete field consists of field points separated by distance $d$. The size of the lattice is finite, $l$. The relation between $l$ and the number of points, $N$, is $l = Nd$. Consider a Lagrangian that contains kinetic energy terms such as $\frac{1}{2}\dot{\phi}_j^2$ and potential energy terms $\frac{g}{2d^2}(\phi_j - \phi_{j-1})^2$ that describe field interaction with nearest neighbor. The strength of the interaction is represented by constant $g$. The field equation for the scalar field $\phi_j(t)$ is

$$\ddot{\phi}_j = \frac{c^2}{d^2}(\phi_{j-1} - 2\phi_j + \phi_{j+1}), \qquad (20)$$

The strength of the interaction, $g$, is related to the wave speed as $c^2 = g$.

A solution for Eq. 20 is given by $\phi_j = ae^{i(kx-\omega t+\delta)}$ where $x = jd$ is position, $a$ is the amplitude and $\delta$ is a phase. Replacing the solution into Eq. 20, we obtain

$$\omega = 2\frac{c}{d}\sin\left(\frac{kd}{2}\right), \qquad (21)$$

We note that applying the long wavelength limit, $k \to 0$, to Eq. 21, we find the standard relation, $\omega = ck$.

Imposing boundary conditions, $\phi_{j=0} = \phi_{j=N+1} = 0$, we get $N$ solutions,

$$\phi_{j;r}(t) = a\sin(k_r dj)\, e^{i\omega_r t}, \qquad (22)$$

where

$$k_r = \frac{\pi}{(N+1)d}r, \qquad (23)$$

and

$$\omega_r = \frac{2c}{d}\sin\left(\frac{r\pi}{2(N+1)}\right), \qquad (24)$$

for $r = 1, 2, \ldots, N$.

The zero-point energy in the gap of length $nd$, between the plates, is

$$U(n) = \frac{\pi\hbar c}{d}\sum_{r=1}^{n}\sin\left(\frac{r\pi}{2(n+1)}\right). \qquad (25)$$

The energy of the system is

$$U_S(n,N) = \frac{\pi\hbar c}{d}\left[\sum_{r=1}^{n}\sin\left(\frac{r\pi/2}{n+1}\right) + \sum_{r=1}^{N-n}\sin\left(\frac{r\pi/2}{N-n+1}\right)\right]. \qquad (26)$$

We note that the sums Eq. 26 may be done exactly using one of Lagrange's trigonometric identities [9], which for large $n$ is

$$\sum_{r=1}^{n}\sin\left(\frac{r\pi}{2(n+1)}\right) = \frac{1}{2}\cot\left(\frac{\pi}{4(n+1)}\right). \qquad (27)$$

The zero-point energy in Eq. 26 is now

$$U_S(n,N) = \frac{\pi\hbar c}{2d}\left[\cot\left(\frac{\pi}{4(n+1)}\right) + \cot\left(\frac{\pi}{4(N+n+1)}\right)\right], \qquad (28)$$

For the range $N \gg n \gg 1$ Eq. 28 gives,

$$\mathcal{F} = -\frac{\pi}{24}\frac{\hbar c}{x^2}. \qquad (29)$$

The force in Eq. 29 is standard value in Eq. 1. It is important to note that to obtain Eq. 29 we do not need to let the lattice spacing $d$ go to zero. We conclude that the Casimir force for a spatially discrete field is indistinguishable from the continuous field result after regularization. We suspect that in higher dimensions, spatially discrete scalar fields would produce the correct Casimir forces. However, we have not found identities, similar to Lagrange's identities in Eq. 27, to simplify the sums. This means that in higher dimensions, we must do the additions numerically but we would only obtain approximate results.

We illustrate numerical approximations to the sums in Eq. 26 using Mathematica. The numerical calculation for the range $N \gg n \gg 1$ gives the constants $A$ and $b$ in formula

$$\mathcal{F} = -\frac{\pi}{A}\frac{\hbar c}{d^{2-b}y^b}. \qquad (30)$$

For instance, when $N = 17\,620$ and we change $n = 600 \to 705$ in steps of 5, we obtain $A = 24.11$, $b = 2.00$. This value is nearly the standard value, $A = 24$, $b = 2$. In general, we find that if the range $N \gg n \gg 1$ is not satisfied, then $b$ is far from 2 and $A$ is far from 24. We also find that whenever $b$ is close to 2 then $A$ is automatically close to 24.

Since using the analytical expression in Eq. 27 for the sum results in the exact force, we might expect that picking larger values for $N > 17\,620$ would improve numerical results but this does not happen. We think that this is due to accumulation of error in larger calculations. For instance, when $N = 90\,000$ and we change $n = 600 \to 705$ in steps of 5, we obtain $A = 24.54$, $b = 1.997$. We find a similar pattern in higher dimensions. Thus, we seek values of $N$ and $n$ that optimize the calculation.

Finally, we find that the Casimir effect in a lattice is easily affected by the chopping the highest energy mode contributions. For instance, when $N = 17620$ and we change $n = 600 \rightarrow 705$ in steps of 5, and we chop 2 parts in a 1000 of the top values of $N$ in Eq. 26, we get $A = 33.53$, $b = 1.945$. When we chop 5 parts in 1000 of the top values of $N$ in Eq. 26, we get $A = 759$, $b = 1.42$. This value is far from the standard value, $A = 24$, $b = 2$.

**Square Lattice**

The simplest structure we can analyze in two dimensions is the square lattice in Fig. 2.

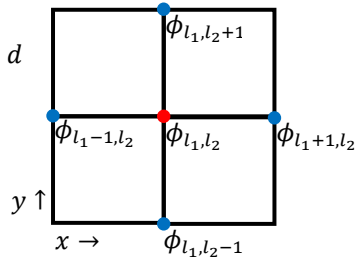

**Fig. 2** The time independent square lattice setup. We use the integers $l_1$ and $l_2$ to represent steps in the $x$ and $y$ directions respectively, with the center field represented by $\phi_{l_1,l_2}$.

The dynamical equation for $\phi_{l_1,l_2}$ is

$$\ddot{\phi}_{l_1,l_2} = \frac{g}{d^2}(-4\phi_{l_1,l_2} + \phi_{l_1,l_2-1} + \phi_{l_1,l_2+1} + \phi_{l_1-1,l_2} + \phi_{l_1+1,l_2}). \tag{31}$$

We assume a plane wave solution to Eq. 31,

$$\phi_{l_1,l_2}(t) = a e^{i(\vec{k}\cdot\vec{l}-\omega t)}, \tag{32}$$

where $a$ is the amplitude, $\vec{k}$ is the wavevector, $\vec{l}$ is the position on a lattice site, $\omega$ is the frequency and $t$ is time.

We introduce lattice vectors

$$\vec{a}_1 = \hat{x}\,d \quad \text{and} \quad \vec{a}_2 = \hat{y}\,d, \tag{33}$$

where $d$ is Planck length. The position of an arbitrary point in is

$$\vec{l} = l_1 \vec{a}_1 + l_2 \vec{a}_2, \tag{34}$$

where $l_1$ and $l_2$ are intergers. The reciprocal lattice vectors are [10]

$$\vec{b}_1 = \hat{x}\,\frac{1}{d} \quad \text{and} \quad \vec{b}_2 = \hat{y}\,\frac{1}{d}. \tag{35}$$

A wavevector is

$$\vec{k} = k_1 \vec{b}_1 + k_2 \vec{b}_2. \tag{36}$$

A property of vectors in Eqs. 33 and 35 is

$$\vec{a}_i \cdot \vec{b}_j = \delta_{ij}. \tag{37}$$

The plane wave solution in Eq. 32 is now

$$\phi_{l_1,l_2} = A e^{i(k_1 l_1 + k_2 l_2 - \omega t)}. \tag{38}$$

In Eq. 38, $k_1$ and $k_2$ refer to wavevectors along the $x$ and $y$ directions respectively. After substituting Eq. 38 in Eq. 31, we find the frequency

$$\omega^2 = 4\frac{c^2}{d^2}\left\{\sin^2\left(\frac{k_1}{2}\right) + \sin^2\left(\frac{k_2}{2}\right)\right\}. \tag{39}$$

where $c = \sqrt{g}$.
We plot the dispersion relation in Eq. 39 in Fig. 3.

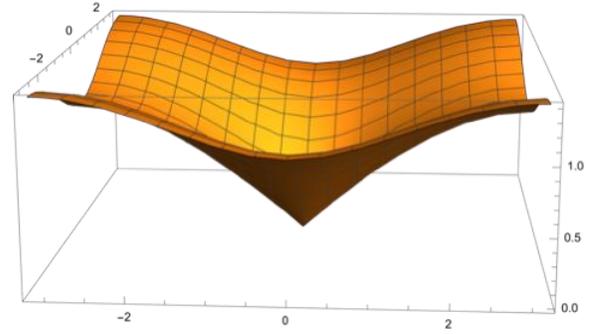

**Fig. 3** Frequency $\omega$ as a function of wavevector $\vec{k}$, for the square lattice. At low frequencies, the dispersion relation has the shape of a cone which indicates rotational symmetry.

For large wavelengths, $\lambda \gg d$, Fig. 3 shows that the dispersion relations given by Eq. 39 resembles a cone,

$$\omega = c\,|\vec{k}|, \tag{40}$$

Equation 40 exhibits rotational symmetry since $\omega$ is independent of the direction of the wavevector $\vec{k}$.

The group velocity that corresponds to dispersion relation in Eq. 39 is defined as $v_g = \frac{d\omega}{dk}$. For wavelengths $\lambda \gg d$ we have $v_g = c$. As the wavelength approaches the shortest wavelength, $\lambda \to 2d$, rotational symmetry is lost; however, the group velocity approaches zero, $v_g \to 0$. Thus, high frequency waves have minor effects and overall rotational symmetry is preserved.

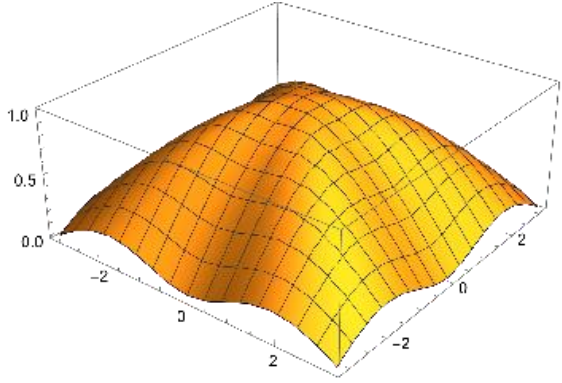

**Fig. 4** Group velocity for the square lattice. At zero frequency, at the center of figure, the group velocity is the highest. At high frequencies, the group velocity goes to zero. At intermediate frequencies, rotational symmetry is partly lost.

We choose periodic boundary conditions along the $x$-axis and fixed boundary conditions along the $y$-axis,

$$\phi_{0,l_2} = \phi_{N_x,l_2}$$
$$\phi_{l_1,0} = \phi_{l_1,N_y+1} = 0, \quad (41)$$

with $N_x$ and $N_y$ representing the size of the lattice in their respective directions. It is easy to check that separation of variables technique works for this case; thus, the time independent part of the field can be written as

$$\phi_{l_1,l_2} = A e^{i k_1 l_1} \sin(k_2 l_2), \quad (42)$$

where $k_1 = \frac{2 r \pi}{N_x}$ for $r = 1,2,\ldots,N_x$, and $k_2 = \frac{s \pi}{N_y+1}$ for $s = 1,2,\ldots,N_y$.

The frequency in Eq. 39 becomes

$$\omega_{r,s} = \frac{2c}{d} \sqrt{\sin^2\left(\frac{r\pi}{N_x}\right) + \sin^2\left(\frac{s\pi}{2(N_y+1)}\right)}, \quad (43)$$

where $r = 1,2,\ldots,N_x$ and $s = 1,2,\ldots,N_y$. We check that Eq. 43 obeys the correct long wave limit, $f\lambda = c$, for periodic waves along the $x$-direction. For the purpose of checking this limit, we ignore wave motion along the $y$-direction and select the smallest non-zero wavevector along the $x$-direction, represented in Eq. 43 by $r = 1$ and $s = 0$,

$$\omega_{1,0} = \frac{2c}{d}\left(\frac{\pi}{N_x}\right). \quad (44)$$

Since the corresponding wavelength should be equal to the width of lattice, $\lambda_{1,0} = N_x d$, Eq. 44 yields the correct long wave limit,

$$f_{1,0} = \frac{c}{\lambda_{1,0}}. \quad (45)$$

For waves with fix ends along the y-direction, the smallest non-zero wavevector in Eq. 43 corresponds to $r = 0$ and $s = 1$. The corresponding wavelength is twice the length of the lattice, $\lambda_{0,1} = 2(N_y + 1)d$. Thus, for this choice, Eq. 43 also gives the correct relation,

$$f_{0,1} = \frac{c}{\lambda_{0,1}}. \quad (46)$$

The Casimir effect calculation is similar the one-dimensional case. We consider the square lattice of size $N_x \times N_y$ in Fig. 5.

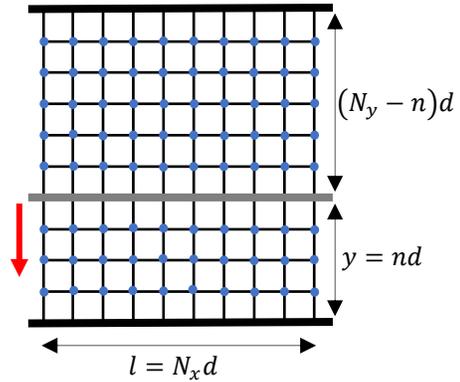

**Fig. 5** The movable barrier located at $y$, divides the lattice into two parts. The barrier experiences an attractive force towards the nearest wall.

The distance $y$ is

$$y = (n+1)d \sim nd. \quad (47)$$

The length of the barrier is

$$l = N_x d. \quad (48)$$

The zero-point energy of the system is

$$U_S(n, N_x, N_y) = \frac{\hbar}{2}\sum_{r=1}^{N_x}\sum_{s=1}^{n}\omega_{r,s} + \frac{\hbar}{2}\sum_{s=1}^{N_x}\sum_{r=1}^{N_y-n}\omega_{r,s} \quad (49)$$

The first term in Eq. 49 is the energy between the plates,

$$\frac{\hbar c}{d}\sum_{r=1}^{N_x}\sum_{s=1}^{n}\sqrt{\sin^2\left(\frac{r\pi}{N_x}\right)+\sin^2\left(\frac{s\pi}{2(n+1)}\right)}, \quad (50)$$

and the second term in Eq. 49 is the energy in the environment,

$$\frac{\hbar c}{d}\sum_{r=1}^{N_x}\sum_{s=1}^{N_y-n}\sqrt{\sin^2\left(\frac{r\pi}{N_x}\right)+\sin^2\left(\frac{s\pi}{2(N_y-n+1)}\right)}. \quad (51)$$

The force is $F_y = -\frac{dU_S}{dy} = -\frac{dU_S}{dn}\frac{dn}{dy}$, where the factor $\frac{dn}{dy} = \frac{1}{d}$ comes from Eq. 47. Dividing the force by the length of the barrier, $l = N_x d$, we obtain the force per unit length,

$$\mathcal{F} = -\frac{1}{A}\frac{\hbar c}{d^{3-b}y^b}. \quad (52)$$

where the constants $A$ and $b$ come from a numerical calculation. When we use $N_x = N_y = 860$ and change $n$ from $150 \rightarrow 190$ in steps of 5 we get

$$\mathcal{F} = -\frac{1}{20.89}\frac{\hbar c}{y^{3.00}}. \quad (53)$$

This result is in close agreement with the standard value in Eq. 1, $A = 20.9$ and $b = 3$. See Appendix.

It has been rightly argued that the Casimir effect in electrodynamics is primarily a low energy phenomenon. Conducting plates reflect low energy photons while allowing high energy photons to pass through. This observation provides a plausible explanation for why the specific type of lattice used in the calculation does not significantly affect the Casimir effect. In the lattices we employ, the dispersion relation is isotropic at low energy, and it is only at high energy that direction becomes significant. As a result, we investigate the impact of truncating a minute fraction of the highest energy contributions to the Casimir effect.

We find that the Casimir effect in 2D is easily affected by the chopping the highest energy mode contributions. When we chop 2 parts in a 1000 to the top values of $N_x$ in Eqs. 50 and 51, we get $A = 0.13$, $b = 4.22$. This value is far from the standard value, $A = 20.9$, $b = 3$. This finding provides compelling evidence that high energy contributions do play a significant role in the lattice model of the Casimir effect.

**Barriers at 45 Degrees**

It is possible to impose boundary conditions with parallel barriers at 45 degrees as seen in Fig. 6. We expect that the Casimir force would be affected by this change.

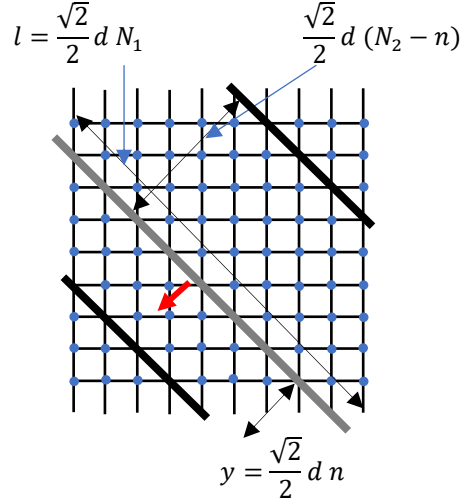

**Fig. 6** A section of the square lattice with fixed barriers at 45°. Fixed barriers are drawn in black. The movable barrier in gray moves along the direction of the red arrow.

To calculate in this lattice, we use vectors in Eqs. 33 and 35. Waves that travel perpendicular to the barrier are described by

$$\vec{k}_\perp = \frac{k_\perp}{\sqrt{2}}(\vec{b}_1 + \vec{b}_2). \quad (54)$$

While waves that travel parallel to the barriers have wavevector

$$\vec{k}_\parallel = \frac{k_\parallel}{\sqrt{2}}(\vec{b}_1 - \vec{b}_2). \quad (55)$$

Assuming that the solution allows separation of variables we write

$$\phi_{l_1,l_2} = a e^{i\vec{k}_\parallel \cdot \vec{l}} \sin(\vec{k}_\perp \cdot \vec{l}) e^{i\omega t}. \quad (56)$$

Equation 56, written explicitly, is

$$\phi_{l_1,l_2} = a e^{ik_\parallel(l_1-l_2)/\sqrt{2}} \sin\left[\frac{k_\perp}{\sqrt{2}}(l_1+l_2)\right] e^{i\omega t}. \quad (57)$$

Inserting the solution in Eq. 57 into Eq. 31 gives the frequency

$$\omega = \frac{2c}{d}\sqrt{1 - \cos\left(\frac{k_\parallel}{\sqrt{2}}\right)\cos\left(\frac{k_\perp}{\sqrt{2}}\right)}. \quad (58)$$

We notice that for small wavevector, Eq. 58 approaches the correct limit, $\omega = c|\vec{k}|$.

We impose fixed boundary conditions along $k_\perp$ and periodic along $k_\parallel$,

$$\phi_{0,l_2} = \phi_{N_1,l_2}$$
$$\phi_{l_1,0} = \phi_{l_1,N_2+1} = 0, \quad (59)$$

with $N_1$ and $N_2$ representing the number of steps in their respective directions. When these boundary conditions are applied to Eq. 57, the allowed frequencies are

$$\omega_{r,s} = \frac{2c}{d}\sqrt{1 - \cos\left(\frac{2\pi r}{N_1}\right)\cos\left(\frac{\pi s}{N_2+1}\right)}, \quad (60)$$

where $r = 1,2,\ldots,N_1$ and $s = 1,2,\ldots,N_2$. We check that Eq. 60 obeys the correct long wave limit, $f\lambda = c$, for periodic waves propagating along the direction parallel to the barriers, represented in Eq. 60 by $r = 1$ and $s = 0$. For this choice, Eq. 60 is

$$2\pi f_{1,0} = \frac{2c}{d\sqrt{2}}\left(\frac{2\pi}{N_1}\right). \quad (61)$$

Using the fact that $f_{1,0} = \frac{c}{\lambda_{1,0}}$, we find the wavelength,

$$\lambda_{1,0} = \left(\frac{\sqrt{2}}{2}d\right) N_1. \quad (62)$$

Since this wavelength should be equal to the width of the lattice, then, Eq. 60 is correct provided that the length of the smallest step along the direction parallel to the barriers is $\left(\frac{\sqrt{2}}{2}d\right)$. Thus, the length of the barrier is

$$l = \left(\frac{\sqrt{2}}{2}d\right) N_1. \quad (63)$$

The smallest wavevector that is perpendicular to the barriers is represented in Eq. 60 by $r = 0$ and $s = 1$. For fix boundary conditions, the wavelength should be twice the lattice length. A similar procedure as in the previous case shows that the corresponding wavelength is

$$\lambda_{0,1} = \left(\frac{\sqrt{2}}{2}d\right) 2(N_2+1). \quad (64)$$

Therefore, the smallest step along the direction perpendicular to the barrier is also $\left(\frac{\sqrt{2}}{2}d\right)$. Thus, distance $y$, in Fig. 4, in terms of smallest step is

$$y = \left(\frac{\sqrt{2}}{2}d\right) n. \quad (65)$$

We note that the size of the smallest cell in this calculation is

$$\left(\frac{\sqrt{2}}{2}d\right)^2 = \frac{1}{2}V. \quad (66)$$

which is ½ the size of the smallest cell for the square lattice, $V = d^2$. There is one field point per smallest cell. Since for the scalar field there is one mode per field point, we have effectively double the number of modes. Considering that the Casimir force is proportional to the number of modes, the force found is twice as strong as it should. The correction is simply to divide the final force by two.

For the calculation, we consider the lattice of size $N_1 \times N_2$. The movable barrier of length $l$ divides the lattice into two parts. The zero-point energy of the system is

$$U_S(n) = \frac{\hbar}{2}\sum_{r=1}^{N_1}\sum_{s=1}^{n}\omega_{r,s} + \frac{\hbar}{2}\sum_{s=1}^{N_1}\sum_{r=1}^{N_2-n}\omega_{r,s}. \quad (67)$$

The first term in Eq. 67 is

$$\frac{\hbar c}{d}\sum_{r=1}^{N_1}\sum_{s=1}^{n}\sqrt{1 - \cos\left(\frac{2\pi r}{N_1}\right)\cos\left(\frac{\pi s}{n+1}\right)} \quad (68)$$

and the second term is

$$\frac{\hbar c}{d}\sum_{r=1}^{N_1}\sum_{s=1}^{N_2-n}\sqrt{1 - \cos\left(\frac{2\pi r}{N_1}\right)\cos\left(\frac{\pi s}{N_2-n+1}\right)}. \quad (69)$$

The force is $F_y = -\frac{dU_S}{dy} = -\frac{dU_S}{dn}\frac{dn}{dy}$, where the factor $\frac{dn}{dy} = \frac{2}{\sqrt{2}d}$ comes from Eq. 65. The force per unit length is obtained by dividing the force by the length of the barrier in Eq. 63 and dividing by two to account for the double counting of modes.

When, in the numerical calculation, we use $N_1 = N_2 = 860$ and change $n$ from $150 \rightarrow 190$ in steps of 5 we get

$$\mathcal{F} = -\frac{1}{20.90}\frac{\hbar c}{y^{3.00}}. \quad (70)$$

A result that is in excellent agreement with the values obtained from Eq. 1, $A = 20.9$ and $b = 3$.

We initially anticipated that the Casimir force in two dimensions would rely on the orientation of the plates with respect to the lattice. However, our calculations have revealed that this is not the case. Even when barriers are positioned diagonally to the natural lattice direction in the square lattice, the Casimir force remains unaltered.

**The Triangular Lattice**

In the triangular lattice, fields are arranged at the vertices of equilateral triangles with vectors $(\vec{a}_1, \vec{a}_2)$ as shown in Fig. 7.

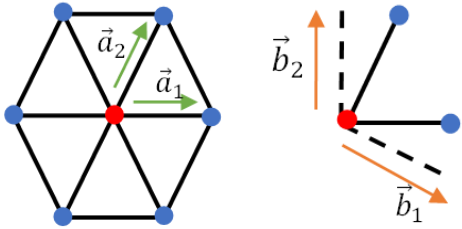

**Fig. 7** Triangular lattice describe by vectors $\vec{a}$ and reciprocal lattice vectors $\vec{b}$. Vector $\vec{a}_1$ runs along the direction of the base of the triangle and vector $\vec{a}_2$ runs along the side of the triangle at 60° from the base.

The lattice vectors are

$$\vec{a}_1 = \hat{x} d$$

$$\vec{a}_2 = \frac{d}{2}\hat{x} + \frac{\sqrt{3}d}{2}\hat{y}. \quad (71)$$

A general vector is given by

$$\vec{l} = l_1 \vec{a}_1 + l_2 \vec{a}_2, \quad (72)$$

where $l_1$ and $l_2$ are integers. The reciprocal lattice vectors [11] in Fig. 5 are

$$\vec{b}_1 = \frac{1}{d}\hat{x} - \frac{1}{\sqrt{3}d}\hat{y}$$

$$\vec{b}_2 = \frac{2}{\sqrt{3}d}\hat{y}. \quad (73)$$

The dynamical equation for field $\phi_{l_1,l_2}$ is

$$\ddot{\phi}_{l_1,l_2} = \frac{g}{d^2}\big(-6\phi_{l_1,l_2} + \phi_{l_1-1,l_2+1} + \phi_{l_1+1,l_2-1} + \phi_{l_1-1,l_2} + \phi_{l_1+1,l_2} + \phi_{l_1,l_2-1} + \phi_{l_1,l_2+1}\big). \quad (74)$$

We first try the plane wave solution

$$\phi_{l_1,l_2} = a e^{i(\vec{k}\cdot\vec{l} - \omega t)}, \quad (75)$$

with wave vector

$$\vec{k} = k_x \frac{\hat{x}}{d} + k_y \frac{\hat{y}}{d}. \quad (76)$$

In terms of reciprocal lattice vectors, Eq. 76 is

$$\vec{k} = k_x\left(\vec{b}_1 + \frac{1}{2}\vec{b}_2\right) + k_y \frac{\sqrt{3}}{2}\vec{b}_2. \quad (77)$$

Using Eqs. 72 and 77, we find that

$$\vec{k}\cdot\vec{l} = k_x l_1 + \frac{1}{2}(k_x + \sqrt{3}k_y)l_2. \quad (78)$$

Replacing Eqs. 75 and 78 into Eq. 74, we obtain

$$\omega = \frac{\sqrt{2g}}{d}[3 - \cos(k_x) - 2\cos\left(\frac{\sqrt{3}}{2}k_y\right)\cos\left(\frac{k_x}{2}\right)]^{1/2}. \quad (79)$$

The dispersion relation in Eq. 79 and corresponding group velocity, $v_g = \frac{d\omega}{dk}$, are plot in Fig. 8. At intermediate frequencies, the dispersion relation and group velocity maintain rotational symmetry much better than the square lattice.

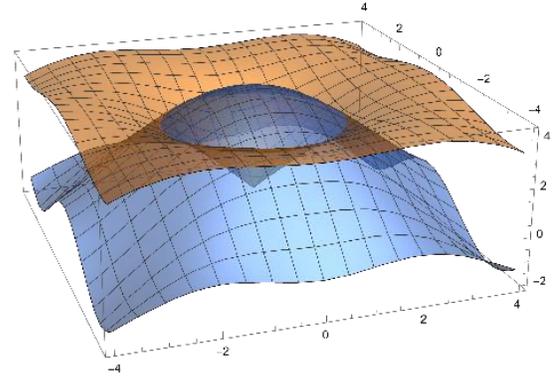

**Fig. 8** The dispersion relation (yellow) and group velocity (blue) for the triangular lattice. At high frequencies, near edges, the group velocity goes to zero. At intermediate frequencies, rotational symmetry is preserved.

The relation in Eq. 79 for small $k$ reduces to the continuous relation,

$$\omega = c\,|\vec{k}|, \quad (80)$$

provided that $c = \sqrt{\frac{3g}{2}}$.

The solution to Eq. 74 that we implement is

$$\phi_{l_1,l_2} = ae^{ik_x\left(l_1+\frac{1}{2}l_2\right)} \sin\left(\frac{\sqrt{3}}{2}k_y l_2\right) e^{i\omega t}. \quad (81)$$

The first factor in Eq. 81, represents periodic waves that propagate along the $x$-direction. The second factor represents standing waves that propagate along the $y$-direction, as seen in Fig. 9.

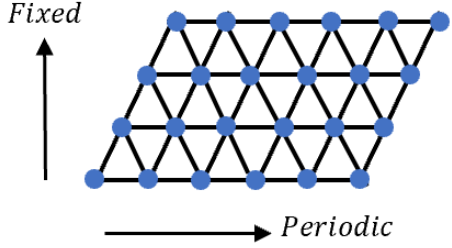

**Fig. 9** Periodic waves propagate along the $x$-axis. The lattice could be thought of as wrapped around a cylinder with its axis along the $y$-direction. The fixed barriers are located at the top and bottom of the cylinder. The movable barrier is parallel to the fixed barriers.

Replacing Eq. 81 into Eq. 74, we find the same frequency is as in Eq. 79. By imposing boundaries conditions to Eq. 81, we obtain

$$k_x = \frac{2\pi r}{N_x} \quad \text{and} \quad k_y = \frac{2s\pi}{\sqrt{3}(N_y+1)}, \quad (82)$$

where $r = 1,2,\ldots N_x$ and $s = 1,2,\ldots N_y$. Equation 79 becomes

$$\omega_{r,s} = \frac{2c}{\sqrt{3}d}\left[3 - \cos\left(\frac{2\pi r}{N_x}\right) - 2\cos\left(\frac{s\pi}{N_y+1}\right)\cos\left(\frac{\pi r}{N_x}\right)\right]^{1/2}. \quad (83)$$

We check that Eq. 83 obeys the correct long wave limit, $f\lambda = c$. The wave with the longest wavelength that propagate along the $x$-direction is represented in Eq. 83 by by $r = 1$ and $s = 0$,

$$2\pi f_{1,0} = \frac{2c}{d}\left(\frac{\pi}{N_x}\right). \quad (84)$$

Using the fact that $f_{1,0} = \frac{c}{\lambda_{1,0}}$, we find that the wavelength is correctly given by

$$\lambda_{1,0} = N_x d. \quad (85)$$

Therefore, the smallest step along the direction parallel to the barriers is $d$. Thus, we have the relation between distance and steps

$$x = nd. \quad (86)$$

The wave with longest wavelength along the $y$-direction is represented in Eq. 83 by $r = 0$ and $s = 1$. A similar procedure as in previous case shows that the corresponding wavelength is

$$\lambda_{0,1} = \left(\frac{\sqrt{3}}{2}d\right) 2(N_y + 1). \quad (87)$$

From Eq. 87 we see that the smallest step along the direction perpendicular to the barrier is $\left(\frac{\sqrt{3}}{2}d\right)$. Thus, the relation between distance and steps is

$$y = \left(\frac{\sqrt{3}}{2}d\right) n. \quad (88)$$

Therefore, the size of the smallest cell in the calculation is

$$d \times \frac{\sqrt{3}}{2}d = \frac{\sqrt{3}}{2}d^2 = V; \quad (89)$$

thus, no correction is needed.

The calculation of the Casimir force for the triangular lattice is similar to the square lattice calculation. When we use $N_x = N_y = 967$ and change $n$ from $150 \to 190$ in steps of 5 we get

$$\mathcal{F} = -\frac{1}{20.91}\frac{\hbar c}{y^{3.00}}. \quad (90)$$

A result that is in excellent agreement with the standard values obtained from Eq. 1, $A = 20.9$ and $b = 3$.

We initially anticipated that the Casimir force in two dimensions would rely on the type of lattice employed. However, our calculations have revealed that this is not the case. Surprisingly, the Casimir force exhibits the same value regardless of whether a square lattice or a triangular lattice is used in 2D.

**Discrete Time in 2D and 3D**

We test whether the fully discretized wave equation produces the correct Casimir effect in 2 and 3 dimensions. The wave equation for a massless scalar field in 2D is

$$\frac{\partial^2 \phi}{c^2 \partial t^2} = \frac{\partial^2 \phi}{\partial x^2} + \frac{\partial^2 \phi}{\partial y^2}. \tag{91}$$

To consider the fully discretized wave equation, we introduce an interval of time given by $t = s\tau$, where $s$ is an integer and $\tau$ is the unit of time that corresponds to Planck length, $\tau = d/c$. For the square lattice case, the equation of motion for $\phi_{l_1, l_2; s}$ is

$$\frac{1}{c^2 \tau^2}(\phi_{l_1,l_2;s+1} + \phi_{l_1,l_2;s-1} - 2\phi_{l_1,l_2;s}) = \frac{1}{d^2}(-4\phi_{l_1,l_2;s} + \phi_{l_1,l_2-1;s} + \phi_{l_1,l_2+1;s} + \phi_{l_1-1,l_2;s} + \phi_{l_1+1,l_2;s}). \tag{92}$$

We substitute in Eq. 92 the plane wave solution

$$\phi_{l_1,l_2;s} = A e^{i(k_x l_1 + k_y l_2 - \omega s)}. \tag{93}$$

The result is

$$\sin^2\left(\frac{\omega}{2}\right) = \sin^2\left(\frac{k_x}{2}\right) + \sin^2\left(\frac{k_y}{2}\right). \tag{94}$$

The Casimir effect is supposed to be a low energy effect; thus, we might only need to consider low energy contributions for $k_x$ and $k_y$. To fulfill the condition in Eq. 94, we must limit the high energy contributions. Due to the symmetry of space, we assume similar high energy range for $k_x \approx k_y \approx k$; thus, Eq. 94 requires that

$$2\sin^2\left(\frac{k}{2}\right) \leq 1 \tag{95}$$

Therefore, the high energy limit for the wavevector is

$$\frac{k_x}{2} = \frac{k_y}{2} = \frac{\pi}{4} \tag{95}$$

Numerically calculations (see Appendix) shows that when the upper limit for the components of the wavevector is $\pi/4$, the Casimir effect cannot be reproduced. According to our calculations, this situation does not change in three dimensions.

**Concluding remarks**

Since the calculation of the Casimir force for a continuous scalar field in 1D diverges, we attempted to perform the calculation by discretizing spacetime. However, our attempt resulted in a null value. Interestingly, when we consider time continuous and a spatially discrete field, we were able to derive the standard value for the Casimir force in 1D. When we moved to 2D and beyond, we found that the fully discretized wave equation does no work either. However, if we only discretize the spatial components of the scalar field, we obtain the correct results. We find that in 3D the situation is similar to 2D.

It would be intriguing to explore other physical effects, beyond the Casimir effect, that could be successfully calculated using a spatially discrete field in a continuous spacetime. We call this approach Hamiltonian lattice theory. Our approach shares similarities with the Kogut-Susskind Hamiltonian formulation of lattice gauge theories, where time is continuous while space is discretized [11]. It is worth noting that this technical approach is currently experiencing renewed interest [12].

Since the lattice spacing, denoted as $d$, is Planck length, $d = 1.6 \times 10^{-35}$ m, then in practical terms, distances will be much larger than $d$. In this limit, the dispersion relations given by Eqs. 39 and 79 resemble the continuous case relation, $\omega = c\,|\vec{k}|$, which exhibits rotational symmetry. In fact, for the square lattice, the group velocity for waves at an energy of $1.24 \times 10^9$ TeV differs from c by only one part in $10^{13}$, making it difficult to detect. It is worth noting that as we approach energies that correspond to the Planck length scale, the rotational symmetry of the dispersion relations is lost for both lattices. However, the corresponding waves tend not to propagate as their group velocity approaches zero, thereby preserving rotational symmetry. Nevertheless, additional symmetries should be analyzed. We point out that our discrete fields technique would still require renormalization.

We note that superluminal or subluminal wave velocity does not necessarily imply a breakdown of relativistic invariance. For instance, we suspect that superluminal wave propagation is present in the violation of Bell's inequalities [13]. Relativistic invariance applies to the dynamics of quantities such as energy-momentum.

**References**

[1] J & Drell, S. D., 1965, Relativistic Quantum Fields, (McGraw-Hill College)

[2] Collins, J., 1984, Renormalization, (Cambridge Univ. Press,)

[3] Casimir, H. B. G., 1948, Proc. K. Ned. Akad. Wet., 51, 793-95

[4] Bressi, G., Carugno, G., Onofrio, R. and Ruoso, G., Measurement of the Casimir force between Parallel Metallic Surfaces, Phys. Rev. Lett., 88 (4) 2002

[5] Chernodub, M. N., Goy, V. A., and Molochkov, A. V., 2016, Casimir effect on the lattice: U(1) gauge theory in two spatial dimensions arXiv:1609.02323

[6] Actor, A., Bender, I., J. Reingruber, the Casimir Effect on a Finite Lattice, arXiv:quant-ph/9908058

[7] J. Ambjorn and S. Wolfram. Ann. Phys. (NY), 147:1, 1983

[8] Cooke, J. H., 1998, the Casimir Force on a Loaded String, AJP 66, 569

[9] Spiegel, M R., and Liu, J., 1998, Mathematical Handbook of Formulas and Tables (McGraw-Hill Companies, Inc), pp 133-34

[10] Ziman, J. M., 1971, Principles of the Theory of Solids, (Cambridge Univ. Press)

[11] Kogut, J. and Susskind, L., 1975. Hamiltonian formulation of Wilson's lattice gauge theories. Phys. Rev. D. 11 (2): 395–408

[12] Zohar, E., and Burrello, M., 2015, Formulation of lattice gauge theories for quantum simulations, Phys. Rev. D 91, 054506

[13] Bell, J. S., Physics 1 (3): 195–200

## Appendix

Mathematica code for discrete field with continuous time

□ Square lattice

```
Clear[k, Nx, Ny, b, c];
Nx = 860;
Ny = 860;
b = 150;
c = 190;
y[k_] :=
  N[ParallelSum[Sqrt[(Sin[r π/(2(k+1))])^2+(Sin[g π/Nx])^2], {r, 1, k},
     {g, 1, Nx}] + ParallelSum[Sqrt[(Sin[s π/(2(Ny-k+1))])^2+
     (Sin[g π/Nx])^2], {s, 1, Ny-k}, {g, 1, Nx}]] / Nx;
data = Table[{k, y[k]}, {k, b, c, 5}];
intfun = Interpolation[data];
□ We linearize the results using the Log function
ParametricPlot[
  {{-Log[intfun'[k]], Log[k]}, {-Log[1/(20.9*k^3)], Log[k]}},
  {k, b, c}, PlotLegends → "Expressions", PlotRange → All]
```

(plot showing -Log[InterpolatingFunction] and -Log[0.0478468995215311/k^3] vs Log[k])

□ We find the slope and intercept of the line

```
data = Table[{Log[k], -Log[intfun'[k]]}, {k, b, c, 5}];
line = Fit[data, {1, k}, k]
```

3.03942 + 3.00405 k

$e^{3.0394159609633356}$

20.893

"the standard value for the power in Casimir effect is 3; here we get 3.00405"

"the standard value for the coefficient in Casimir effect is 20.9; here we get 20.893"

Mathematica code for field in discrete spacetime

□ chopping the high frequency components so as to avoid complex angular frequencies. The maximum allowed wave-vector components $k_x=k_y$=Pi/2 so as to get a real frequency.

```
Clear[k, Nx, Ny, b, c, a, d];
Nx = 860;
Ny = 860;
b = 150;
c = 190;
a = 4.001;
d = 2.;
y[k_] :=
  N[ParallelSum[ArcSin[Sqrt[(Sin[r π/(2(k+1))])^2+(Sin[g π/Nx])^2]/1.],
     {r, 1, IntegerPart[k/d]}, {g, 1, IntegerPart[Nx/a]}] + ParallelSum[
     ArcSin[Sqrt[(Sin[s π/(2(Ny-k+1))])^2+(Sin[g π/Nx])^2]/1.],
     {s, 1, IntegerPart[(Ny-k)/d]}, {g, 1, IntegerPart[Nx/a]}]] / Nx;
data = Table[{k, y[k]}, {k, b, c, 5}];
intfun = Interpolation[data];
ParametricPlot[{{-Log[intfun'[k]], Log[k]}, {-Log[1/(20.9*k^3)], Log[k]}},
  {k, b, c}, PlotLegends → "Expressions", PlotRange → All]
```

(plot)

```
data = Table[{Log[k], -Log[intfun'[k]]}, {k, b, c, 5}];
line = Fit[data, {1, k}, k]
```

(1.78087 - 10.6052 i) + (0.249346 + 1.7941 i) k